\documentclass[prd,twocolumn,nofootinbib]{revtex4-1}
\usepackage{graphicx, epsfig}
\usepackage[dvipsnames]{xcolor}
\usepackage{hyperref}
\usepackage{mathrsfs}
\usepackage{amsmath}
\usepackage{enumerate}
\usepackage{bm}
\usepackage{float}
\usepackage{ulem}

\newcommand{\be}{\begin{equation}}
\newcommand{\ee}{\end{equation}}
\newcommand{\bea}{\begin{eqnarray}}
\newcommand{\eea}{\end{eqnarray}}
\newcommand{\ba}{\begin{eqnarray}}
\newcommand{\ea}{\end{eqnarray}}

\newcommand{\gapp}{\mathrel{\raise.3ex\hbox{$>$}\mkern-14mu
              \lower0.6ex\hbox{$\sim$}}}
\newcommand{\lapp}{\mathrel{\raise.3ex\hbox{$<$}\mkern-14mu
              \lower0.6ex\hbox{$\sim$}}}

\begin{document}
\title{Vacuum Topology and the Electroweak Phase Transition}
\author{Yiyang Zhang$^\dag$, Francesc Ferrer$^\dag$, Tanmay Vachaspati$^*$}
\affiliation{
$^\dag$Physics Department and McDonnell Center for the Space Sciences, 
Washington University, St. Louis, MO 63130, USA.\\
$^*$ Physics Department, Arizona State University, Tempe, AZ 85287, USA.\\
$^*$ Maryland Center for Fundamental Physics, University of Maryland,
                    College Park, Maryland 20742, USA.
}

\begin{abstract}
\noindent
We investigate if the topology of pure gauge fields in the electroweak vacuum can play a role in
classical dynamics at the electroweak phase transition. 
Our numerical analysis shows that magnetic fields are produced if the initial vacuum has non-trivial 
Chern-Simons number, and the fields are helical if the Chern-Simons number {\it changes} during the 
phase transition.
\end{abstract}

\maketitle

An explanation for the observed cosmic matter-antimatter asymmetry likely requires
CP violating particle interactions at energies at or above the electroweak scale,
at an epoch when the universe was out of thermal equilibrium~\cite{sakharov1967violation}. 
Models of matter-genesis, more specifically, baryogenesis or leptogenesis, also necessarily involve the 
violation of baryon plus lepton (B+L) number through anomalous quantum processes. Several studies 
have now shown that the anomalous violation of B+L at the time of 
electroweak symmetry breaking, when the Higgs ($\Phi$) acquires a non-vanishing vacuum expectation 
value (VEV), leads to the production of helical magnetic 
fields~\cite{cornwall1997speculations,vachaspati2001estimate}. 
 The connection of matter-genesis and magneto-genesis offers a means to probe fundamental particle 
 interactions by the observation of magnetic fields in the universe.

In hindsight it is not difficult to intuitively understand the production of helical 
magnetic fields when B+L is violated by anomalous processes. To change B+L, requires 
a change in the Chern-Simons number of the electroweak gauge fields and,
post electroweak symmetry breaking, this requires passage through a ``sphaleron''
\cite{Manton:1983nd}
that has the interpretation of a twisted magnetic monopole-antimonopole configuration \cite{vachaspati1994electroweak,hindmarsh1994origin,Saurabh:2017ryg}
The decay of the sphaleron corresponds to the annihilation of the monopole and antimonopole, with
the release of helical magnetic fields \cite{copi2008helical,chu2011magnetic}.

In the present paper, we address a related question -- can the topology of the electroweak
vacuum play a role in the dynamics of the electroweak phase transition?
A hint that the answer is in the affirmative is suggested by the work of 
Jackiw and Pi~\cite{jackiw2000creation},
where they consider a pure vacuum SU(2) gauge field configuration that has non-vanishing
Chern-Simons number. 
They then project the gauge field configuration onto a fixed isospin direction 
to simulate the effects of the Higgs field VEV, and then evolve and calculate the helicity in 
the electromagnetic (EM) field. Jackiw and Pi find a non-vanishing EM helicity and
further provide the neat result that the EM helicity at late times is 1/2 of
the helicity at early times.

As originally discussed in Ref.~\cite{jackiw2000creation}, the Jackiw-Pi result depends 
crucially on their model for projection of the gauge fields in isospin space.
For example, note that the initial gauge configuration is {\it pure} gauge and has zero
energy, while the final configuration with helical magnetic fields has non-zero energy. Clearly
energy is introduced by the act of projecting the non-Abelian gauge fields to the Abelian
component, and it is assumed that the projection somehow mimics electroweak symmetry 
breaking. In a more realistic setting, the projection will be achieved by the process by which 
the Higgs field acquires a VEV, and the precise projection in isospin space
depends on the dynamics of the Higgs field as it interacts with the gauge (and other)
fields.

\begin{figure}
  \includegraphics[height=0.35\textwidth,angle=0]{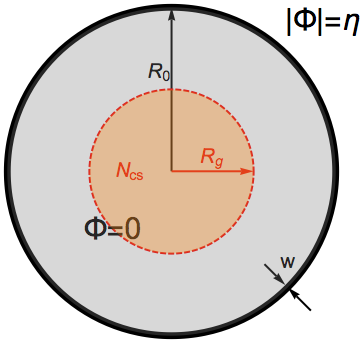}
  \caption{
  In a first-order electroweak phase transition, bubbles of true vacuum
  ($|\Phi| = \eta$) will grow and encapsulate regions of false vacuum 
  ($\Phi=0$), 
 within which gauge fields with localized non-trivial Chern-Simons number, $N_{\rm{CS}}$, 
   may exist. With time,
  the bubbles of true vacuum will grow and complete the phase transition, forcing a
  projection of the gauge fields 
  within the region of radius $R_g$ (shown as an orange disk)
  in the false vacuum onto the EM field. 
  In studying this process,
  we will replace the complicated geometry of the symmetric phase by a spherical
  bubble of radius $R_0$ (thick solid circle) and thickness $w$. 
  }
\label{setup1}
\end{figure}

In this paper we will resolve the effect of isospin projection on the gauge fields by
studying the full dynamics of the Higgs field and the electroweak gauge fields. 
Our first analysis corresponds to dynamics during a first order electroweak phase 
transition. We will
set up a pure gauge field configuration in a spherical region with vanishing Higgs VEV,
surrounded by the true vacuum of the model where the Higgs has already acquired a 
VEV as shown in Fig.~\ref{setup1}. As the spherical region with vanishing Higgs
shrinks, the electroweak gauge fields will get projected on to the EM field
and presumably some magnetic field will be generated. We calculate the energy and helicity 
of the magnetic field as a function of time, for several different values of the initial 
Chern-Simons number.  

\begin{figure}
  \includegraphics[height=0.3\textwidth,angle=0]{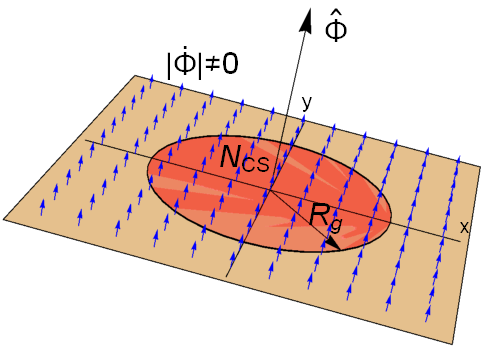}
  \caption{
	 As a second-order electroweak 
	  phase transition proceeds, $\Phi=0$
  everywhere but ${\dot \Phi}\ne 0$. (This is a simplification of the second
  order phase transition dynamics but one that is sufficient for our purposes.) 	
  The drawing shows a spatial slice over which $\Phi=0$ and with a localized region 
	of pure gauge field with non-vanishing Chern-Simons number 
	(red region). The blue arrows in the vertical direction 
	illustrate the initial growth of the Higgs field VEV 
	as in Eq.~(\ref{2nd-order-pt}). As the VEV of $\Phi$ grows, 
	pure gauge field configurations get projected into massless EM fields 
	and massive gauge fields. 
	  }
\label{setup2}
\end{figure}

We also examine the case of a second order electroweak phase transition, as
shown in Fig.~\ref{setup2}. Here the Higgs VEV vanishes
everywhere at the initial time but its time derivative is non-vanishing.

We start in Sec.~\ref{sec:model} by describing the electroweak model and the initial conditions 
that we will use to study the evolution. We describe our numerical results in Sec.~\ref{sec:results}
both for a first-order transition (Sec.~\ref{firstorder}) and for a second-order transition 
(Sec.~\ref{secondorder}). We conclude in Sec.~\ref{sec:conclusions}.
Further details of our numerical setup are provided in the appendices.

\section{Model details}
\label{sec:model}

The bosonic electroweak variables are the Higgs field $\Phi$, the SU(2) valued gauge
fields $W^a_\mu$ and the U(1) hypercharge gauge field $B_\mu$, with Lagrangian
\begin{equation}
\mathcal{L}=\vert D_\mu \Phi\vert^2-\frac{1}{4} W^a_{\mu\nu} W^{a\mu\nu} 
                    - \frac{1}{4} B_{\mu\nu} B^{\mu\nu} -\lambda (\vert \Phi \vert^2 -\eta^2)^2,
\label{lagrangian}
\end{equation}
where
\be
D_\mu = \partial_\mu - i \frac{g}{2} \sigma^a W^a_\mu - i \frac{g'}{2} B_\mu
\ee
is the covariant derivative,
$\sigma^a$ ($a=1,2,3$) the Pauli spin matrices,
and $W^a_{\mu \nu}$, $B_{\mu \nu}$ are the field strengths.
We describe the resulting equations of motion and our numerical techniques to solve them
in Appendices~\ref{sec:equations} and \ref{sec:numerical}.

The initial conditions for the gauge fields are always pure gauge but can have
non-trivial Chern-Simons number. As described in~\cite{jackiw2000creation}, these are
\begin{equation}
W_\mu \equiv \frac{ \sigma^a}{2i}W^a_{\mu} = \frac{1}{g}U^{-1} \partial_\mu U \ , \ \ B_\mu =0,
\label{pure-gauge}
\end{equation}
where, 
\begin{equation}
U = \cos\frac{f(r)}{2} +i\bm{\sigma}\cdot \hat{\bm{\omega}} 
	\sin \frac{f(r)}{2},
\label{su2-mat}
\end{equation}
and the unit vector $\hat{\bm{\omega}}={\hat x}$ points in the radial direction
and $r=|{\bm x}|$ is the radial spherical coordinate.
The function $f(r)$ is chosen to be
\begin{equation}
f(r) = 2\pi n \tanh(r/R_g), \ \ n=0,1,2,\ldots
\label{fr}
\end{equation}
Therefore the profile function $f(r)$  satisfies the boundary conditions $f(0)=0$ and $f(\infty)=2\pi n$. 
The time derivatives of all gauge fields are taken to vanish at the initial time $t=0$. 

The Chern-Simons number is defined as
\begin{eqnarray}
\nonumber N_{\rm CS}(t) &=& \frac{N_F}{32\pi^2}\epsilon^{ijk}\int d^3x \biggl [-g'^2B_{ij}B_k \\ \
&& +g^2 \left ( W^a_{ij} W^a_k -\frac{g}{3} \epsilon_{abc} W^a_i W^b_j W^c_k \right ) \biggr ],
\label{cs}
\end{eqnarray}
where $N_F$ is the number of
fermion families. In the rest of the paper, we will choose $N_F=1$. 
The initial gauge field configuration described in 
Eq.~(\ref{pure-gauge}) has a Chern-Simons number
\begin{equation}
N_{\rm CS}= - n.
\end{equation}

For the first-order phase transition set up shown in Fig.~\ref{setup1}, the Higgs doublet 
is given by
\begin{eqnarray}
\Phi (t=0,{\bm r}) &=&  \frac{\eta}{2} \left [ 1+\tanh \left ( \frac{r-R_0}{w}\right ) \right ] 
\begin{pmatrix} 0 \\ 1 \end{pmatrix} \notag \\
\dot\Phi(t=0,{\bm r})&=&0,
\label{1st-order-pt}
\end{eqnarray}
where $R_0 > R_g$ is the initial size of the false vacuum region (the bubble), 
and $w \ll R_0$ is the width of the transition region from true to false
vacuum, {\it i.e.} the bubble wall thickness.

One question that arises in the setup of the first order phase transition is that it should
be possible to study the evolution after performing a large gauge transformation that
makes the gauge fields trivial, $W_i=0$. If there is exactly zero overlap between the scalar
and gauge profiles, such a gauge transformation will not affect the Higgs field (up to
an overall sign). However, for profile functions that are analytic, such as the ones in
Eqs.~(\ref{fr}) and (\ref{1st-order-pt}), there is always some overlap between the gauge 
fields and non-zero Higgs VEV. In this region $|D_i\Phi|\ne 0$ even for $i$ in the angular 
directions. The large gauge transformation that sets $W_i=0$ will also twist the Higgs.
In fact, the electroweak model has two distinct and independent winding 
numbers: the gauge winding of Eq.~(\ref{cs}) and the Higgs winding, $N_w$, {\it e.g.} 
as defined in~\cite{Mou:2017zwe}, and the difference of the two windings is invariant
even under large gauge transformations. By gauging away the gauge winding, we will 
induce a corresponding Higgs winding. We have explicitly checked that the overlap 
between the Higgs and the gauge fields plays a crucial role in the evolution by also 
considering {\it non-analytic} profile functions, {\it i.e.} where the interior of the bubble 
has exactly $\Phi=0$ and the exterior of the gauge configuration has exactly $W_i=0$. 
Such non-analytic profiles are not expected to be relevant in a physical setting but they
do show that no magnetic energy is produced if there is no overlap.
The importance of the overlap will also be seen for analytic profiles 
when we demonstrate that the magnetic field energy grows with larger $R_g$.

Whereas for a second-order phase transition, the Higgs doublet is 
initially taken to be
\begin{eqnarray}
\Phi(t=0,{\bm r}) &=& 0 \notag \\
\dot\Phi(t=0,{\bm r}) &=& \gamma \, \eta^2
\begin{pmatrix} 0 \\ 1 \end{pmatrix}, 
\label{2nd-order-pt}
\end{eqnarray}
where $\gamma$ is a dimensionless parameter denoting the speed with which
the uniform Higgs is rolling off the top of the potential.
Our modelling of the second order phase transition is not completely
accurate, since we hold the Higgs field at the origin until the potential has
reached its zero temperature form. A more realistic treatment would take 
into account the temperature evolution of the potential over time scales
set by the Hubble expansion, which is $\sim 10^{17}$ times slower than the 
electroweak time scale that determines the dynamical evolution rate of the 
fields in our simulations. We leave a more detailed investigation of these
effects for future work. Nevertheless, we expect our treatment to fit 
more closely the actual cosmological phase transition than the sudden 
projection of the Chern-Simons vacuum
onto massive and massless gauge field components that was used in
Ref.~\cite{jackiw2000creation}.

Once the Higgs has left the symmetric phase, $\Phi=0$,
we can track the EM magnetic field. 
The EM field potential is generally defined by
\begin{equation}
A_\mu = \sin \theta_w n^a W^a_\mu + \cos \theta_w B_\mu,
\label{usualAmu}
\end{equation}
and the EM field strength follows the definition in \cite{vachaspati1991magnetic},
\begin{eqnarray}
A_{\mu\nu}&=&\sin \theta_w n^a W^a_{\mu\nu}+\cos\theta_w B_{\mu\nu} \nonumber \\
&& \hskip -1.1 cm
-i \frac{2}{g\vert\Phi\vert^2} \sin\theta_w
[ (D_\mu \Phi )^\dagger (D_\nu \Phi ) - (D_\nu \Phi)^\dagger (D_\mu \Phi)],
\label{usualAmunu}
\end{eqnarray}
where, $\theta_w$ is the Weinberg angle, and $n^a$ is the unit vector in SU(2)
isospace defined by the direction of the Higgs field 
\begin{equation}
n^a = -\frac{\Phi^\dagger \sigma^a \Phi}{\vert \Phi \vert ^2}.
\end{equation}
These expressions are only defined when $| \Phi | \ne 0$. We shall alter them slightly so that the
definition makes sense for all $\Phi$ and coincides with the usual definition in the symmetry
broken phase. The expressions we use are
\begin{align}
	&A_{\mu\hphantom{\nu}} = \sin \theta_w N^a W^a_\mu + \cos \theta_w B_\mu \label{em-field} \\
	& \notag \\
	&A_{\mu\nu}=\sin \theta_w N^a W^a_{\mu\nu}+\cos\theta_w B_{\mu\nu} \notag\\
&  \quad	-i \frac{2}{g\eta^2} \sin\theta_w
 [ (D_\mu \Phi )^\dagger (D_\nu \Phi ) - (D_\nu \Phi)^\dagger (D_\mu \Phi)],
 \label{em-strength}
\end{align}
where, 
\begin{equation}
N^a = -\frac{\Phi^\dagger \sigma^a \Phi}{\eta^2}.
\end{equation}

We will also calculate the magnetic energy in the EM field
\begin{equation}
E_{\rm em} = \frac{1}{2} \int d^3x ~ {\bm B}^2,
\label{mag-energy}
\end{equation}
and the magnetic helicity
\begin{equation}
H_{\rm em} = \int d^3x ~ {\bm A}\cdot {\bm B}.
\label{helicity}
\end{equation}

Our numerical scheme is based on a lattice implementation of the electroweak
equations as described in \cite{rajantie2001electroweak}. The numerical details are listed 
in Appendix \ref{sec:numerical}. 
We adopt phenomenological values of all parameters:
$g=0.65$, $\sin^2\theta_w=0.22$, $g'=g\tan\theta_w$, $\eta=1$, $\lambda=0.129$. 
The Higgs mass is $m_H=2\eta \sqrt{\lambda}=125~\text{GeV}$, therefore $\eta=174~\text{GeV}$, which means 
one unit of energy in our simulation is equivalent to $174~\text{GeV}$. One unit of length is 
$1.13\times 10^{-16} ~\text{cm}$, and one unit of time is $3.78\times10^{-27}~\text{s}$.

We use absorbing boundary conditions (ABC) to minimize effects from lattice 
boundaries and to ensure that negligible contributions enter from outside
the finite lattice box. It should be noted that the specific form of 
the ABC varies, depending on the initial conditions,
as described in Appendix~\ref{abc}. 
We run our simulation as long as the gauge fields are confined within the 
lattice box. 
Conservation of total energy and fulfillment of Gauss constraints are two 
non-trivial checks that we monitor in the simulation. The total energy
is conserved within $1\%$, while the Gauss constraints given 
in~Eq.~(\ref{gauss-hamiltonian}) are satisfied to an even higher accuracy.
Notice that both sets of initial conditions considered in this paper
automatically satisfy the Gauss constraints, and thus should be preserved 
during the evolution in the bulk of the lattice; there may be small violations due
to the boundary conditions on the lattice as discussed in Appendix~\ref{sec:numerical}.

As a final check of our code, we have compared some results with a completely separate evolution code
\cite{Vachaspati:2016abz} and obtained consistent results.

\section{Results}
\label{sec:results}

We will now describe the results of our simulations, first for the first order phase transition set up
of Fig.~\ref{setup1}, and then for the second order phase transition set up of Fig.~\ref{setup2}.

\subsection{First order phase transition}
\label{firstorder}

\begin{figure}
\includegraphics[height=0.29\textwidth,angle=0]{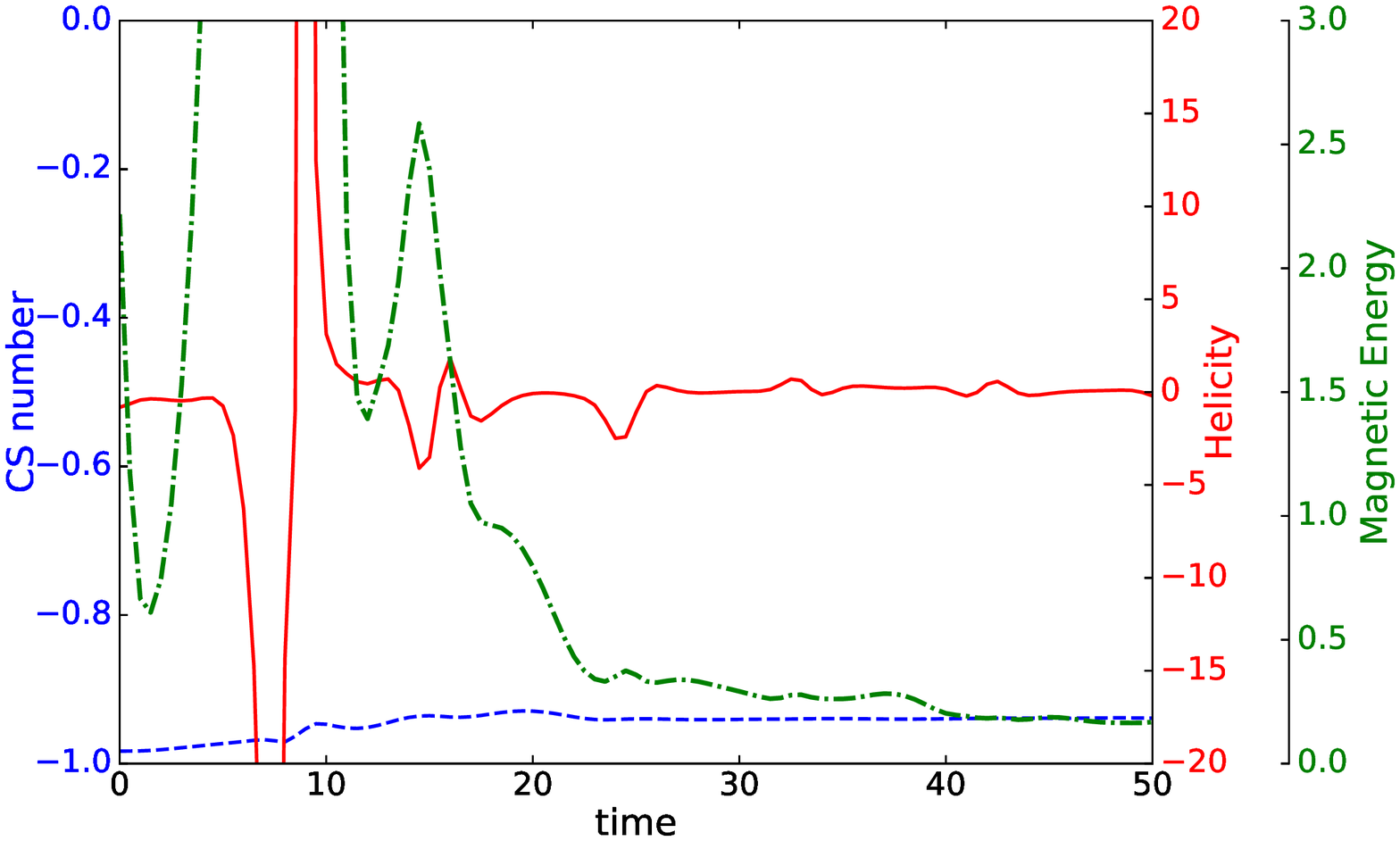}
\includegraphics[height=0.29\textwidth,angle=0]{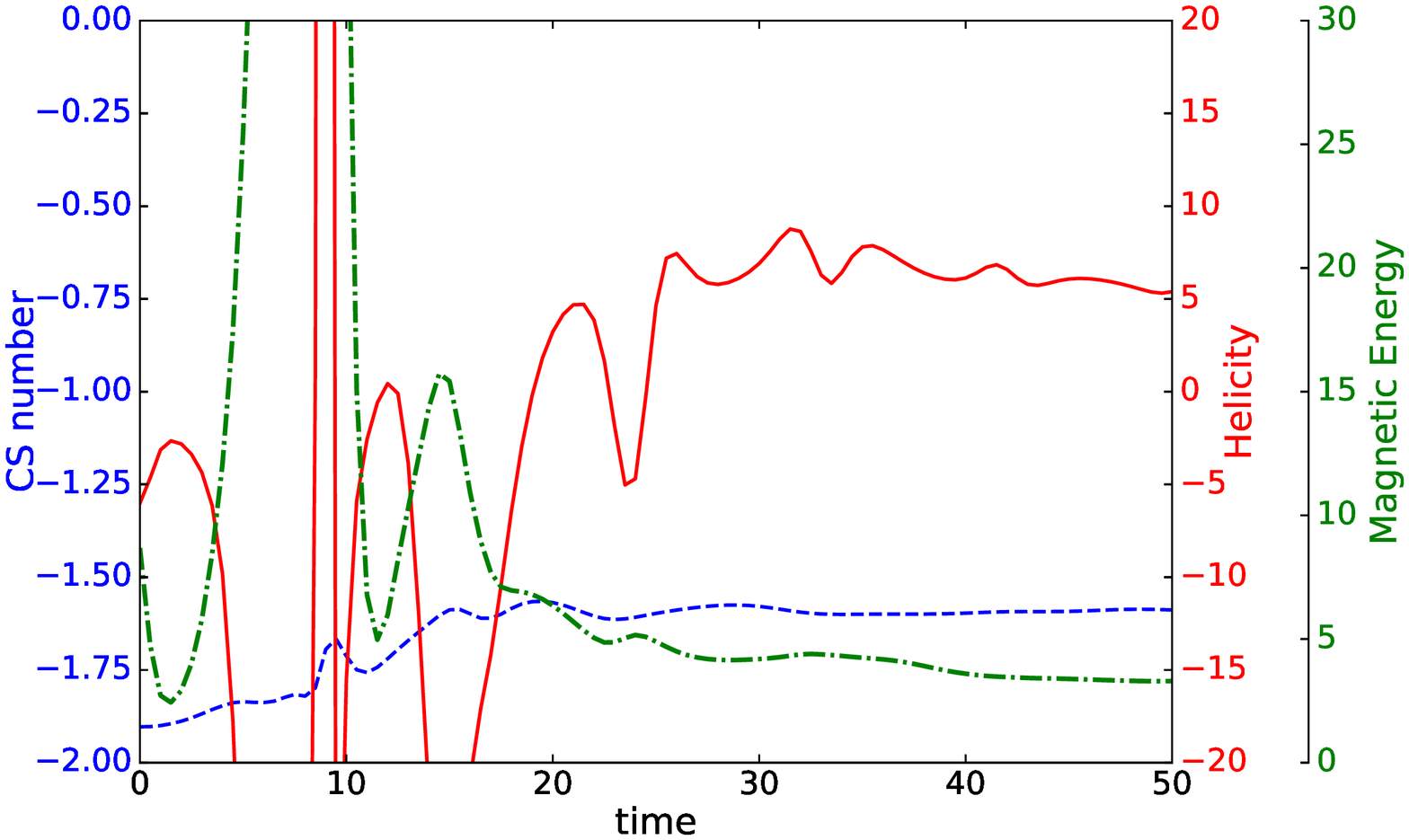}
\includegraphics[height=0.29\textwidth,angle=0]{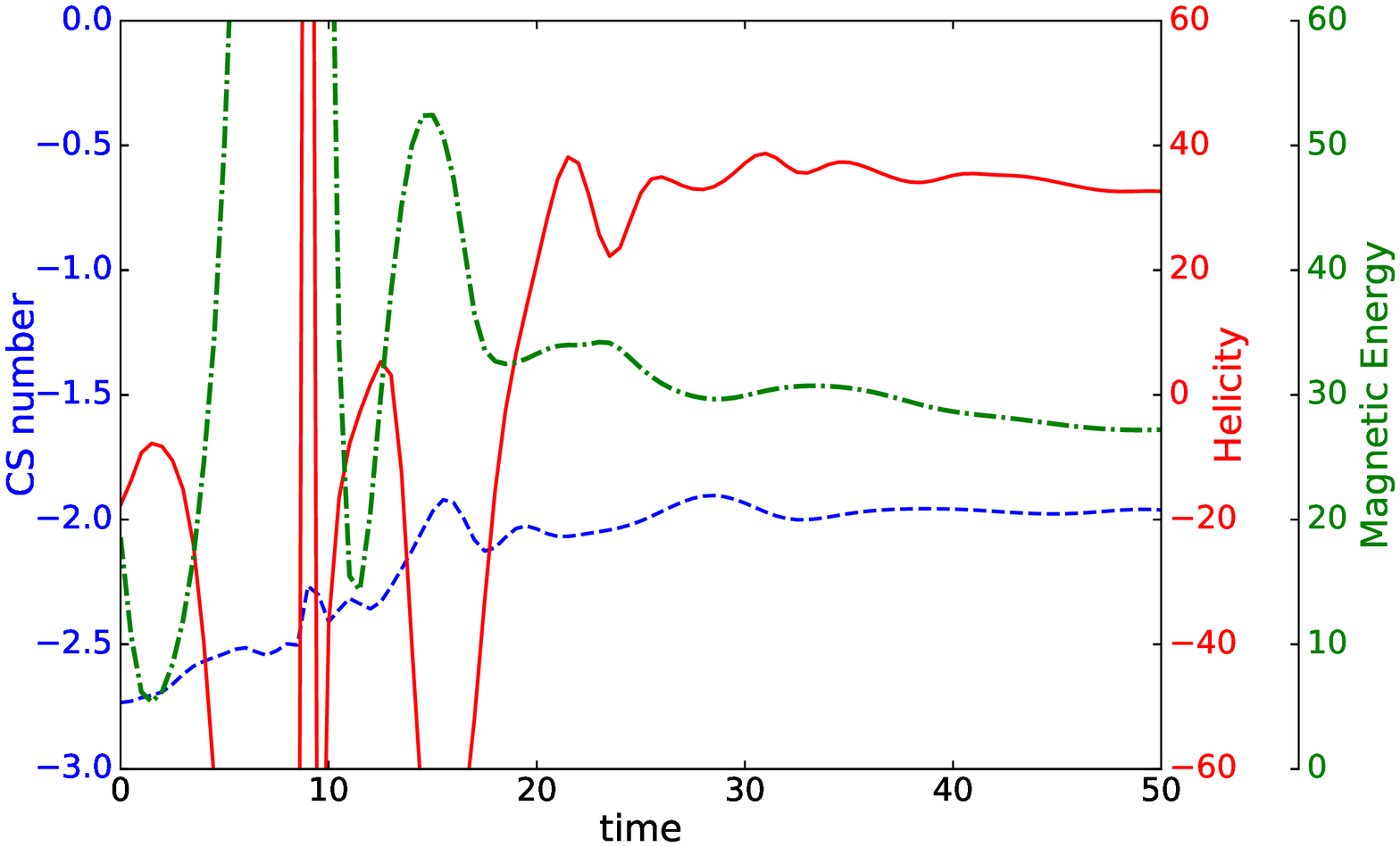}
  \caption{Plots of the Chern-Simons number (blue-dashed),
  magnetic helicity (red solid curve), and magnetic energy (green dot-dashed curve)
  for initial Chern-Simons number $n=1$ (top), $n=2$ (middle), and $n=3$ (bottom)
  in the first-order phase transition case.
	Note the different scale bars for the EM energy (in lattice units)
	shown on the right-hand side of the plot.
	}
\label{case-a-pic}
\end{figure}

The Higgs field configuration at the initial time is given by Eq.~(\ref{1st-order-pt}).
For our simulations we will set the false vacuum bubble radius to be $R_0=8.0$ and the
bubble wall width to be $w=0.4$. At the center of the bubble we start with a pure gauge 
configuration as given in Eq.~(\ref{pure-gauge}) with radius $R_g=6.0$. 
We denote the initial Chern-Simons number by $-n$. The case with $n=0$ 
has trivial evolution and no gauge fields are excited by the bubble collapse. The results for 
$n=1,2,3$ are non-trivial and are shown in Fig.~\ref{case-a-pic} where we plot the evolution
of the Chern-Simons number, the EM magnetic helicity defined in Eq.~(\ref{helicity})
and the EM magnetic energy defined in Eq.~(\ref{mag-energy}). We have tested
the evolution with the definition of EM given in Eqs.~(\ref{usualAmu}) and (\ref{usualAmunu}) 
and find agreement at late times where these expressions are well-defined.

The total energies in our three runs with $n=1,2,3$ are about 964, 1083 and 1250, respectively
and are well above the sphaleron barrier, which is $E_{\rm sph} \approx 9~{\rm TeV}$ 
\cite{Manton:1983nd,Klinkhamer:1984di}
or about 52 in lattice units. So there is ample energy in the simulations for 
the Chern-Simons number
to change. Nevertheless, being above the sphaleron energy barrier 
is a necessary but not sufficient condition for the change in Chern-Simons 
number. 
Also note that the Chern-Simons number is an integer 
only for the vacuum; a non-vacuum configuration may have a non-integer value of the Chern-Simons 
number. Indeed, the plots in Fig.~\ref{case-a-pic} 
show non-integer values of the Chern-Simons number as we always
have some energy in the lattice.

In all our runs, there is some energy transferred from the false vacuum bubble to the 
gauge sector during evolution. This is seen in the plots of the EM energy, which is
non-vanishing after the collapse of the bubble is complete. A non-vanishing positive helicity 
is obtained for the $n=2,3$ cases but not for the $n=1$ case for which the Chern-Simons
number remains roughly constant. A rough fit to the data in Table~I gives
\be
| H_{\rm em} | \approx 56 (\Delta N_{\rm CS})^2
\label{hfit}
\ee
where $\Delta N_{\rm CS}$ is the change in Chern-Simons number,
and $n$ is the initial Chern-Simons number.
Further, the sign of magnetic helicity is the same as the sign
of the change in Chern-Simons number.

\begin{table}[H]
\begin{ruledtabular}
\begin{tabular}{ r  l  c }
$n$ & $\Delta N_{\rm CS}$  & $H_{\rm em}$\tabularnewline
\hline 
1 & $0.044$ & $-0.20$\tabularnewline
\hline 
2 & $0.31$  & $5.4$\tabularnewline
\hline 
3 & $0.77$  & $33$\tabularnewline
\end{tabular}
\end{ruledtabular}
\caption{ $n$, $\Delta N_{\rm CS} $ and $H_{\rm em}$ for the first-order phase transition simulations.
For $n=1$, $\Delta N_{\rm CS}$ and $H_{\rm em}$ are consistent with zero.}
\end{table}

\begin{figure}
	\includegraphics[height=0.3\textwidth,angle=0]{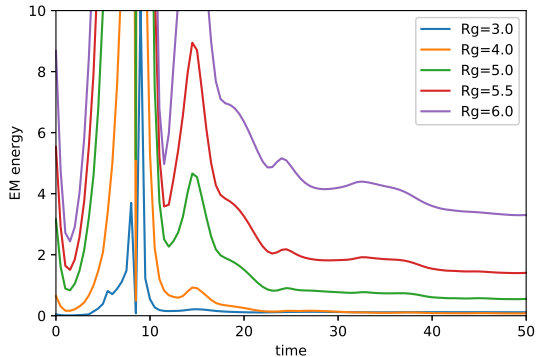}
	\caption{Plot of EM energy generated during a first-order phase transition
		for different $R_g$. The initial conditions are given by Eq. (\ref{pure-gauge}) 
		and Eq. (\ref{1st-order-pt}). Here, $R_0=8.0$, $w=0.4$, $n=2$. }
	\label{em_rg}
\end{figure}

In Fig.~\ref{em_rg} we show the electromagnetic energy as a function of time for several
different values of the initial gauge radius $R_g$. The electromagnetic energy is larger for 
larger $R_g$. This is consistent with our expectations as discussed below Eq.~(\ref{1st-order-pt})
since larger $R_g$ provides greater overlap of the initial gauge fields and the
imploding bubble.

\subsection{Second order phase transition}
\label{secondorder}

The Higgs is now initially assumed to be in the symmetry unbroken phase everywhere
but in the process of rolling down the potential toward the true vacuum (see Fig.~\ref{setup2}
and Eq.~(\ref{2nd-order-pt})).
In this case we need to specify the initial velocity of the Higgs field, the extent of the gauge
field configuration, and the initial Chern-Simons number. We set the velocity parameter
$\gamma=0.4$ and the radius of the gauge field configuration $R_g=6.0$.
Three different initial Chern-Simons number are considered: $n=1,2,3$. As in the first-order
phase transition case, the evolution for $n=0$ is trivial.

\begin{figure}
\includegraphics[height=0.29\textwidth,angle=0]{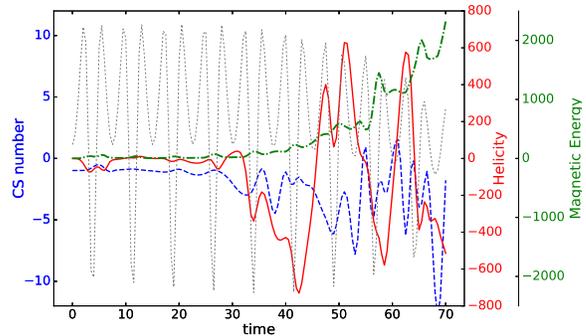}
\includegraphics[height=0.29\textwidth,angle=0]{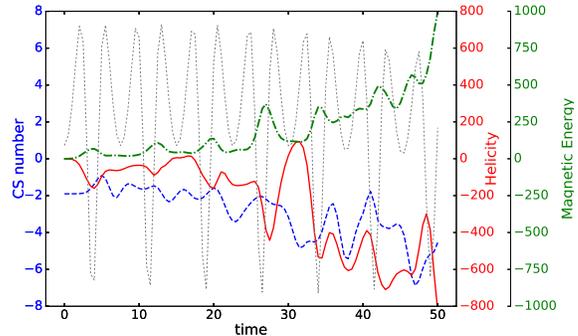}
\includegraphics[height=0.29\textwidth,angle=0]{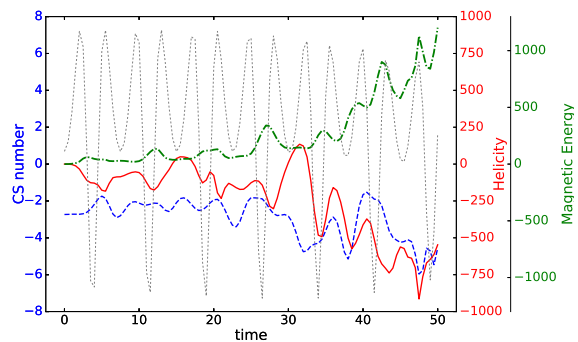}
  \caption{Plots of the Chern-Simons number (blue-dashed),
  magnetic helicity (red solid curve), and magnetic energy (green dot-dashed curve)
  for initial Chern-Simons number $n=1$ (top), $n=2$ (middle), and $n=3$ (bottom)
  in the second-order phase transition case. The kinetic energy of the Higgs is
  also shown in grey.
  Note the different scale bar for the EM energy shown in lattice units
  on the right-hand side of the plot.
}
\label{case-d-pic}
\end{figure}

The results for the second-order phase transition are very different from the results of
the first-order phase transition. The oscillatory features can be understood by realizing
that the Higgs field oscillates about the true minimum, as is clear
from the oscillations in the Higgs kinetic energy curve in Fig.~\ref{case-d-pic}. We see
the general feature that Chern-Simons number, magnetic helicity, and magnetic
energy, all grow at late times, when the kinetic energy of the Higgs also starts
dissipating. We can also understand the oscillatory behavior 
by noting that the Higgs field oscillates in the potential
(light grey curve in in Fig.~\ref{case-d-pic}).
The growth of the Chern-Simons number and magnetic helicity and energy,
is more rapid for larger values of $n$. We expect the growth to saturate once the
energy is evenly distributed between the scalar and gauge field sectors. However,
to see this would require very long run times and very large lattices. 

Our results show that even in the case of a second order phase transition, energy
is transferred during the phase transition to EM magnetic fields with non-trivial
helicity if the gauge fields initially have non-vanishing Chern-Simons number.
As in the first-order phase transition case, the sign of magnetic helicity is the 
same as the sign of the change in Chern-Simons number.

\section{Conclusions}
\label{sec:conclusions}

We have investigated the effects of pure gauge fields and their topology on the 
electroweak phase transition.
The creation of magnetic fields from the gauge vacuum was anticipated in 
Ref.~\cite{jackiw2000creation}, treating the phase
transition as a mathematical projection of the gauge fields into massive and massless
components. In this paper we have numerically examined the classical dynamics of
the electroweak symmetry breaking for different gauge vacua. Our results broadly agree 
with the analysis of Ref.~\cite{jackiw2000creation} in that the evolution can lead to the creation 
of helical magnetic fields. 

The details of the evolution are more involved. 
In the present work, we have explored the evolution and magnetic field generation during
processes that can occur during a first-order phase transition that proceeds by bubble 
nucleation, and also during a second-order phase transition that proceeds by a continuously 
rolling Higgs field. In cases where the initial Chern-Simons number is zero and we set the 
gauge fields to zero, the evolution is trivial and the Chern-Simons number continues to be zero 
and magnetic fields are not produced.
If the initial Chern-Simons number is non-zero but does not change during evolution (see the $n=1$ 
case in Fig.~\ref{setup1}), even then helicity is not generated though magnetic fields are
produced. This may be related to the generation of magnetic fields at the electroweak phase 
transition due to non-vanishing gradient energy of the Higgs as was discussed in 
Ref.~\cite{Vachaspati:1991nm,Vachaspati:1994xc}. 
(For this reason we also expect magnetic fields to be produced in the zero Chern-Simons 
case if initially the gauge fields are not zero.)
If the initial Chern-Simons number is 
large -- greater than equal to 2 for the other parameters in our runs -- the Chern-Simons number 
changes during evolution and magnetic helicity is produced. The connection between changes
in the Chern-Simons number and magnetic helicity production, and the relation
with changes in the baryon number via a quantum anomaly, 
has been pointed out in Refs.~\cite{Cornwall:1997ms,Vachaspati:2001nb,copi2008helical}. Thus there 
are several lines of reasoning that point to the production of magnetic fields at the electroweak phase 
transition.

Based on early arguments we would expect that the magnetic helicity is directly
proportional to the change in Chern-Simons number \cite{Cornwall:1997ms,Vachaspati:2001nb}. 
In our runs, this simple relationship does not bear out. Instead we observe that the magnetic 
helicity goes as the square of the change in Chern-Simons number (Eq.~(\ref{hfit})). 
This result should be considered
tentative because we have only been able to run our simulations for a few
values of the initial Chern-Simons number. Finer lattices and longer runs
will be necessary to study a greater range of initial Chern-Simons number.
Quantitative estimates of the magnetic field produced at the electroweak phase
transition will require further work along the lines of \cite{Mou:2017zwe}. 

Finally we also mention that a non-Abelian vacuum consisting of a periodic
array of pure gauge vortices has been proposed in Ref.~\cite{Olesen:2016pxv}. 
It is argued that spontaneous symmetry breaking in such a vacuum can also 
generate magnetic fields~\cite{Olesen:2017mcu,Olesen:2017ykg}. It would be 
worthwhile to study a dynamical phase transition in this vacuum just as we have 
done for the Chern-Simons vacua in this paper.

\acknowledgements
We thank Kohei Kamada, Poul Olesen, Paul Saffin and Anders Tranberg for feedback.
YZ thanks the MCFP, University of Maryland for hospitality.
The computations were done on the A2C2 Saguaro Cluster at ASU and on the 
HPC Center at Washington University.
This work is supported by the U.S. Department of Energy, Office of High Energy Physics, 
at ASU under  Award No. DE-SC0013605 at ASU and at Washington University
under Award No. DE-FG02-91ER40628.

\appendix

\section{Electroweak continuum equations}
\label{sec:equations}

The classical electroweak equations of motion that result 
from the bosonic electroweak Lagrangian
in~Eq.(\ref{lagrangian}) are: 
\begin{align*}
	&D_\mu D^\mu \Phi + 2\lambda (\vert\Phi\vert^2-\eta^2)\Phi=0 \\ 
	&\partial_\mu B^{\mu\nu} = g' \text{Im} \big[ \Phi^\dagger (D^\nu \Phi) \big] \\ 
	&\partial_\mu W^{a\mu\nu} + g\epsilon^{abc} W^b_\mu W^{c\mu\nu} = g\text{Im} \big[ \Phi^\dag \sigma^a(D^\nu\Phi) \big].
\end{align*}

In a numerical simulation, it is convenient to use the temporal gauge, 
$W_0^a=0$ and $B_0=0$. Then, the equations of motion become:
\begin{align}
	\partial_0^2 \Phi\hphantom{{}_i} =& D_i D_i \Phi - 
	2\lambda (\vert \Phi \vert ^2 -\eta^2) \Phi  \notag\\ 
	\partial_0^2 W^a_i =&-\partial_k W^a_{ik}  \notag \\ 
&- g\epsilon^{abc}W^b_k W^c_{ik} + g\text{Im}[\Phi^\dagger \sigma^a(D_i\Phi)] 
	\notag \\  
\partial_0^2 B_i =& -\partial_k B_{ik}+g'\text{Im}[\Phi^\dagger(D_i\Phi)],
\end{align}
along with two Gauss constraints,
\begin{align}
	&\partial_0 \partial_j B_j -g'\text{Im}\big[ \Phi^\dagger \partial_0 \Phi \big] = 0 \notag \\ 
	& \partial_0 \partial_j W^a_j + g \epsilon^{abc} W^b_j \partial_0 W^c_j 
        -g\text{Im} \big[ \Phi^\dagger \sigma^a \partial_0 \Phi \big] =0.
\label{gauss-2}
\end{align}

We have implemented a discretized version of these equations following
Ref.~\cite{Vachaspati:2016abz} as a check of our main results that 
were obtained using the lattice formulation of Appendix~\ref{sec:numerical}.

\section{Lattice implementation}
\label{sec:numerical}

Our lattice implementation of the electroweak evolution equations 
follows closely the 
discussion in Ref.~\cite{rajantie2001electroweak}. 

We introduce the lattice based fields $U_\mu(t,x)$ and $V_\mu(t,x)$, which 
are related to the continuum gauge fields through:
\ba
U_i(t,x)&=& \exp{\Big(-\frac{i}{2} g\Delta x \sigma^a W^a_i\Big)}\notag \\
U_0(t,x)&=& \exp{\Big(-\frac{i}{2} g\Delta t \sigma^a W^a_0\Big)}\notag  \\
V_i(t,x)&=& \exp{\Big(-\frac{i}{2} g\Delta x B_i\Big)}\notag  \\
V_0(t,x)&=& \exp{\Big(-\frac{i}{2} g\Delta t B_0\Big)}. 
	\label{gaugelatt}
\ea

The discretized action in terms of these fields is:
\begin{widetext}
\begin{eqnarray}
\nonumber
S &=& \sum_{x,t} \Delta t \Delta x^3 \Bigg\{ \big(D_0\Phi \big)^\dagger \big(D_0\Phi \big) 
- \sum_i \big(D_i\Phi \big)^\dagger \big(D_i\Phi \big) 
-U(\Phi) + \Big(\frac{2}{g\Delta t \Delta x}\Big)^2\sum_i \Big(1-\frac{1}{2}\text{Tr}~U_{0i}\Big) \\ 
&&+\Big(\frac{2}{g'\Delta t \Delta x}\Big)^2 \sum_i \Big(1-\text{Re} ~V_{0i}\Big) 
-\frac{2}{g^2 \Delta x^4} \sum_{i,j} \Big(1-\frac{1}{2} \text{Tr}~U_{ij}\Big) 
	- \frac{2}{g'^2\Delta x^4} \sum_{i,j} \Big(1-\text{Re}~V_{ij}\Big) \Bigg\},
	\label{actionlattice}
\end{eqnarray}
\end{widetext}
where, $\Phi(t,x)$ is the Higgs field doublet defined on each lattice site; 
$U_i(t,x)$ and $V_i(t,x)$ 
are the SU(2) and U(1) link fields, respectively, 
defined on the link between the neighboring sites $x$ and $x+i$;
and for the U(1) link field we take the real part rather than the trace as in 
Ref.~\cite{rajantie2001electroweak}. Also, $U_0(t,x)=I_2$ and $V_0(t,x)=1$
consistent with~(\ref{gaugelatt}) and our choice of temporal gauge.
Note that throughout this appendix Latin indices take values 
$i,j,k = 1,2,3$ and repeated indices are not summed over.

Here, we adopt the conventional interpretation that $U_i(t,x)$, $V_i(t,x)$ 
parallel transport the fields at site $x+i$ back to site $x$;  then
$U_i^\dagger(t,x)$, $V_i^\dagger(t,x)$ parallel transport the fields at 
site $x$ to $x+i$. Then, the covariant derivative of $\Phi(t,x)$ that
enters in Eq.~(\ref{actionlattice}) has components:
\begin{align*}
	D_i \Phi &= \frac{1}{\Delta x} \big[ U_i(t,x) V_i(t,x) \Phi(t,x+i) - \Phi(t,x)\big] \\
	D_0 \Phi &= \frac{1}{\Delta t} \big[ U_0(t,x) V_0(t,x) \Phi(t+\Delta t,x) -\Phi(t,x) \big].
\end{align*}

Finally, the plaquette fields can be seen as the discretized version of 
the magnetic fields: 
\begin{align*}
U_{ij}(t,x) = U_j(t,x) U_i(t,x+j) U_j^\dagger (t,x+i) U_i^\dagger (t,x) \\
V_{ij}(t,x) = V_j(t,x) V_i(t,x+j) V_j^\dagger (t,x+i) V_i^\dagger (t,x).
\end{align*}

The fields $\Phi(t,x)$, $U_i(t,x)$ and $V_i(t,x)$ are defined at the time steps 
$t+\Delta t$, $t+2\Delta t$, $\ldots$; while the conjugate momentum fields, 
$\Pi(t+\Delta t/2,x)$, $F(t+\Delta t/2,x)$ and $E(t+\Delta t/2,x)$, are 
defined at time steps $t+\Delta t/2$, $t+3\Delta t/2$, $\ldots$. They are related by
\begin{align}
\Phi(t+\Delta t,x) =& \Phi(t,x) +\Delta t \Pi(t+\Delta t/2,x) \\
V_i(t+\Delta t,x) =& \frac{1}{2} g'\Delta x \Delta t E_i(t+\Delta t/2,x) V_i(t,x) \\
U_i(t+\Delta t,x) =& g\Delta x \Delta t F_i(t+\Delta t/2,x) U_i(t,x), 
\end{align}
The equations of motion that result from setting the functional derivative
of the action to zero are:
\begin{widetext}
	\begin{align}
\Pi(t+\Delta t/2,x) =& \Pi(t-\Delta t/2,x)+\Delta t \Big\{ \frac{1}{\Delta x^2}\sum_i \big[ U_i(t,x)V_i(t,x)\Phi(t,x+i) \notag \\
	&-2\Phi(t,x)+U_i^\dagger(t,x-i) V_i^\dagger(t,x-i) \Phi(t,x-i)\big] - \frac{\partial U}{\partial \Phi^\dagger}\Big\}
\label{eom-pi}
		\\ 
\text{Im} [E_k(t+\Delta t/2,x)] =& \text{Im} [E_k(t-\Delta t/2,x)]+\Delta t \Big\{ \frac{g'}{\Delta x}\text{Im}[\Phi^\dagger(t,x+k)U^\dagger_k(t,x) V^\dagger_k(t,x) \Phi(t,x)] \notag \\
&-\frac{2}{g'\Delta x^3}\sum_i \text{Im} [V_k(t,x) V_i(t,x+k) V_k^\dagger (t,x+i) V_i^\dagger (t,x) \notag \\ 
&+ V_i(t,x-i) V_k(t,x) V_i^\dagger(t,x+k-i)V_k^\dagger(t,x-i)] \Big\} \\ 
\text{Tr} [i\sigma^m F_k(t+\Delta t/2,x)] =& \text{Tr} [i\sigma^m F_k(t-\Delta t/2,x)] +\Delta t\Big\{ \frac{g}{\Delta x} \text{Re} [\Phi^\dagger(t,x+k)U_k^\dagger(t,x)V_k^\dagger(t,x) i \sigma^m \Phi(t,x)] \notag \\
&-\frac{1}{g\Delta x^3}\sum_i \text{Tr} [ i\sigma^m U_k(t,x) U_i(t,x+k) U_k^\dagger(t,x+i) U_i^\dagger(t,x) \notag\\
&+i\sigma^m U_k(t,x) U_i^\dagger(t,x+k-i) U_k^\dagger(t,x-i) U_i(t,x-i) ]
		\Big\},
\end{align}
where, in Eq.~(\ref{eom-pi}) the term 
	$U_i^\dagger(t,x-i) V_i^\dagger(t,x-i) \Phi(t,x-i)$ corrects the 
	corresponding equation~(A17) in 
	Ref.~\cite{rajantie2001electroweak} where it is written without the $-i$.
\end{widetext}
The remaining components, $\text{Re} (E_k)$ and $\text{Tr} (F_k)$, can be 
found by using:
\begin{equation}
|E|=\frac{2}{g'\Delta x \Delta t}, \quad  \text{det}(F)=\Big( \frac{1}{g\Delta x \Delta t} \Big)^2,
\end{equation}
where the square in the second equation
corrects a typo in Ref.~\cite{rajantie2001electroweak}.

The lattice action~(\ref{actionlattice}) is invariant under 
gauge transformations: 
\begin{align*}
\Phi(t,x) &\rightarrow \Omega_1(t,x) \Omega_2(t,x) \Phi(t,x) \\ 
U_i(t,x) &\rightarrow \Omega_2(t,x) U_i(t,x) \Omega_2^\dagger (t,x+i) \\
V_i(t,x) &\rightarrow \Omega_1(t,x) V_i(t,x) \Omega_1^\dagger (t,x+i) \\
\Omega_2 &\in SU(2), \quad \Omega_1 \in U(1),
\end{align*}
which imply the following Gauss constraints:
\begin{widetext}
\begin{eqnarray}
\label{gauss-u1}
G_{\rm U1} (x) \equiv \frac{1}{\Delta x} \sum_i \text{Im} [E_i(t+\Delta t/2,x) - E_i(t+\Delta t/2,x-i)] - g' \text{Im} [\Pi^\dagger(t+\Delta t/2,x)\Phi(t,x)]&=&0 \\ 
\label{gauss-su2} \nonumber
	G_{\rm SU2}^k(x) \equiv \frac{1}{\Delta x} \sum_i \text{Tr} ~\{ i\sigma^k \big[F_i (t+\Delta t/2,x)- U_i^\dagger(t,x-i) F_i(t+\Delta t/2,x-i) U_i(t,x-i)\big] \} \\
	- g\text{Re} [\Pi^\dagger(t+\Delta t/2,x)i\sigma^k \Phi(t,x)] &=& 0 
\end{eqnarray}
\end{widetext}
If the initial time is $t_0=0$, our initial conditions should specify 
$\Phi(0,x)$, $U_i(0,x)$, $V_i(0,x)$, as well as $\Pi(\Delta t/2,x)$, 
$F_i(\Delta t/2,x)$, $E_i(\Delta t/2,x)$. This is consistent with the 
requirements for second-order differential equations. As mentioned 
in \cite{moore1996improved}, there is no conserved quantity that can be 
identified as energy, but we can construct a quantity that approaches the 
conserved energy in the small $\Delta t$ limit:
\begin{widetext}
\begin{eqnarray}
\nonumber
E &=& \sum_x \Delta x^3 \Bigg\{ \Big(\frac{\Pi(t+\Delta t/2,x)+\Pi(t-\Delta t/2,x)}{2}\Big)^\dagger \Big(\frac{\Pi(t+\Delta t/2,x)+\Pi(t-\Delta t/2,x)}{2}\Big)  \\ \nonumber
&&+ \sum_i \big[D_i\Phi(t,x) \big]^\dagger \big[D_i\Phi(t,x) \big] +U(\Phi(t,x))  \\ \nonumber
&&+ \Big(\frac{2}{g\Delta t \Delta x}\Big)^2 \frac{1}{2}\sum_i \Big[\Big(1-\frac{g\Delta x\Delta t}{2}\text{Tr}~F_i(t+\Delta t/2,x)\Big)+\Big(1-\frac{g\Delta x\Delta t}{2}\text{Tr}~F_i(t-\Delta t/2,x)\Big) \Big] \\ \nonumber
&&+\Big(\frac{2}{g'\Delta t \Delta x}\Big)^2 \frac{1}{2}\sum_i \Big[\Big(1-\frac{g'\Delta x\Delta t}{2}\text{Re} ~E_i(t+\Delta t/2,x)\Big) + \Big(1-\frac{g'\Delta x\Delta t}{2}\text{Re} ~E_i(t-\Delta t/2,x)\Big)\Big]\\ 
&& +\frac{2}{g^2 \Delta x^4} \sum_{i,j} \Big(1-\frac{1}{2} \text{Tr}~U_{ij}(t,x)\Big)  + \frac{2}{g'^2\Delta x^4} \sum_{i,j} \Big(1-\text{Re}~V_{ij}(t,x)\Big) \Bigg\}
\end{eqnarray}
\end{widetext}

There are two checks that can be made to ensure the simulation is running 
correctly. The first one is conservation 
of total energy. Given a localized configuration, the total energy inside the 
lattice box should be fixed before this configuration reaches the boundary. 
The second check is that the Gauss constraints should be satisfied.
Following Ref.~\cite{rajantie2001electroweak}, we introduce a ``Hamiltonian''
\begin{equation}
\label{gauss-hamiltonian}
H = \frac{(\Delta x)^3}{2}\sum_{x} [G_{\rm U1}(x)G_{\rm U1}(x)+G_{\rm SU2}^k(x)G_{\rm SU2}^k(x)] 
\end{equation}
as a measure of the violation of the Gauss constraints. Numerically, the value of Eq.~(\ref{gauss-hamiltonian}) should 
be very close to zero.
As pointed out in \cite{moore1996improved}, the Gauss constraints are 
preserved by the evolution algorithm as long as they are satisfied by
the initial conditions. However, we are using ABC to evolve the fields 
at the boundaries. The ABC equations, discussed in the following appendix, 
do not in general preserve the Gauss constraints.
We find that our simulations satisfy the constraints to a very high
accuracy, which is a non-trivial check that the boundary conditions are 
appropriate.

\section{Absorbing Boundary Conditions} \label{abc}

To implement absorbing boundary conditions in our simulations, we 
extend the results 
in~\cite{engquist1977absorbing, szeftel2006nonlinear, feng1999absorbing}
as described below. 

Neglecting for the moment interactions with gauge fields, the equation
of motion for the Higgs can be written as
\begin{equation}
[\partial_t^2-\nabla^2+J(\Phi)]\Phi=0,
\label{scalar-wave}
\end{equation}
where $J(\Phi)=2\lambda(\vert \Phi \vert^2-\eta^2)$.

We can formally decompose Eq.(\ref{scalar-wave}) at a boundary as
\begin{equation}
\big[\bm{n}\cdot\nabla-\sqrt{\partial_t^2-\partial_\perp^2+J(\Phi)}\big]\big[\bm{n}\cdot\nabla+\sqrt{\partial_t^2-\partial_\perp^2+J(\Phi)}\big]\Phi=0,
\end{equation}
where $\bm{n}$ is the outward pointing unit normal vector of the boundary,
and $\partial_\perp \equiv \nabla - \bm{n} (\bm{n} \cdot \nabla)$.

To prevent exterior waves from entering the simulation lattice,
while allowing outgoing waves to leave the box, we
require
\begin{equation}
\big[\bm{n}\cdot\nabla+\sqrt{\partial_t^2-\partial_\perp^2+J(\Phi)}\big]\Phi=0.
\label{exact-abc}
\end{equation}
To find a local approximate form of~Eq.(\ref{exact-abc}) that is suitable
for numerical implementation, we need to expand the square root in a power
series.

If $\partial_t^2$ is the dominant term, then 
$-\partial_\perp^2+J(\Phi)$ can be treated as a small perturbation. 
This approximation will be most accurate for waves that hit the boundary
perpendicularly, with negligible $\partial_\perp^2 \Phi$ 
and negligible $J(\Phi)$.
Keeping terms that are linear in the perturbation, 
Eq.(\ref{exact-abc}) becomes 
\begin{equation}
\Bigg[ \bm{n}\cdot\nabla + \partial_t \bigg(1-\frac{\partial_\perp^2-J(\Phi)}{2\partial_t^2} \bigg) \Bigg] \Phi=0,
\end{equation}
which can be simplified as 
\begin{equation}
\partial_t^2 \Phi = -\bm{n}\cdot\nabla \partial_t \Phi +\frac{1}{2} \partial_\perp^2 \Phi -\frac{1}{2}J(\Phi)\Phi.
\label{approx-1st}
\end{equation}
We use this approximation when simulating a first-order phase transition, 
as $J(\Phi)$ can then be neglected at the boundary. 

In the case of a second-order phase transition, 
even for $t=0$ the potential is not 
small at the boundary. Therefore we treat $\partial_t^2+J(\Phi)$ as the 
dominant term and $ \partial_\perp^2 $ as a perturbation. 
Now Eq.(\ref{exact-abc}) becomes
\begin{equation}
\Bigg[ \bm{n}\cdot\nabla +\sqrt{\partial_t^2+J(\Phi)} \bigg( 1-\frac{\partial_\perp^2}{2(\partial_t^2+J(\Phi))}\bigg) \Bigg] \Phi=0
\end{equation}
Assuming that $\partial_t^2 \gg J(\Phi) \gg \partial_\perp^2$, 
we can further simplify the above equation as
\begin{equation}
\partial_t^2 \Phi = -\bm{n}\cdot\nabla \partial_t \Phi +
	\frac{1}{2}\partial_\perp^2 \Phi -J(\Phi)\Phi.
\label{approx-2nd}
\end{equation}

Although this scheme can be extended to include higher order terms in the 
perturbation, we find that Eqs.~(\ref{approx-1st})~and~(\ref{approx-2nd}) are 
accurate enough for our purpose.

To take into account the coupling of the Higgs to the gauge fields,
the spatial derivatives in Eq.(\ref{scalar-wave}) should be replaced with 
covariant derivatives. We start by writing Eq.(\ref{scalar-wave}) as 
\begin{equation}
\label{approx-cov}
[\partial_t^2-\nabla^2+\nabla^2-D^2 +J(\Phi)]\Phi=0, 
\end{equation}
where $D$ denotes the covariant derivative. Then the previous discussion
still holds if we use the current
$J'(\Phi)\equiv \nabla^2-D^2 +J(\Phi)$, which now includes the gauge
interactions.

To evolve the gauge fields at the boundary, we use the lowest order
absorbing boundary conditions:
\begin{equation}
\label{gauge-abc}
\bm{E}^T=-\bm{n}\times\bm{B}
\end{equation}
In principle, higher order corrections could be included along similar lines 
as for the scalar wave equation. However, it has been argued  
in~\cite{feng1999absorbing} that they may give rise to numerical instabilities.

We stress that implementing the ABC for gauge fields on the lattice
requires us to make some compromises that cannot always be rigorously 
justified. We use a scheme that only requires two time slices and the
nearest neighboring spatial points, which roughly speaking is robust as
long as the gauge fields on the boundary are small. As shown by the energy 
and Gauss constraint checks, they are adequate for our purposes.

\bibliographystyle{apsrev}
\bibliography{csToB}

\end{document}